\documentclass[smallextended]{svjour3}

\journalname{epj}
\usepackage{graphicx}

\usepackage{amsmath}
\usepackage[normalem]{ulem}
\usepackage{soul}
\usepackage[utf8]{inputenc}
\usepackage[english]{babel}
\usepackage{lipsum}
\usepackage{csquotes}
\usepackage{amssymb}

\usepackage{marginnote}
\usepackage{cooltooltips}

\usepackage{pstricks}
\newpsobject{showgrid}{psgrid}{subgriddiv=2,griddots=10,gridlabels=3mm}

\begin{document}

\title{Epidemiological impact of waning immunization on a vaccinated population}

%\author{Ewa Grela}
%\author{Michael Stich}
%\author{Amit K Chattopadhyay}

\author{Ewa Grela \and Michael Stich
\and Amit K Chattopadhyay}

\institute{Systems Analytics Research Institute, Mathematics, School of Engineering and Applied Science, Aston University, Aston Triangle, Birmingham, B4 7ET, UK}

\date{}

%\begin{abstract}

\abstract{
This is an epidemiological SIRV model based study that is designed to analyze the impact of vaccination in containing infection spread, in a 4-tiered population compartment comprised of susceptible, infected, recovered and vaccinated agents. While many models assume a lifelong protection through vaccination, we focus on the impact of waning immunization due to conversion of vaccinated and recovered agents back to susceptible ones. Two asymptotic states exist, the  \enquote{disease-free equilibrium} and the \enquote{endemic equilibrium}; we express the transitions between these states as function of the vaccination and conversion rates using the basic reproduction number as a descriptor. We find that the vaccination of newborns and adults have different consequences in controlling epidemics. We also find that a decaying disease protection within the recovered sub-population is not sufficient to trigger an epidemic at the linear level. Our simulations focus on parameter sets that could model a disease with waning immunization like pertussis. For a diffusively coupled population, a transition to the endemic state can be initiated via the propagation of a traveling infection wave, described successfully within a Fisher-Kolmogorov framework. 
}

%\end{abstract}

\maketitle

\section{Introduction}

Infectious diseases have a strong impact on the dynamics of human populations and are routinely highlighted when epidemic outbreaks of deadly infections like Ebola or MERS occur. Increased human mobility, the rise of pathogens resistant to antibiotics (\enquote{Antimicrobial Resistance}), and the advent of new, so called Emerging Infectious Diseases are making infectious diseases a major health challenge of the 21st century. 

To analyze transmitting diseases and epidemic outbreaks and to inform public health organisms, a wide range of mathematical models have been proposed and studied in detail. One of the simplest and well-known models illustrating the dynamics of epidemics is the SIR model, proposed by Kermack and McKendrick in 1927~\cite{KermackPRSA27}. The central idea of this model is to divide the entire population into three separate groups: {\em{susceptible}}~($S$) individuals that have never been infected and are not immune to the disease; {\em infected} ($I$) individuals, who are infectious and can spread the disease within the population; and {\em recovered} (or {\em removed}; $R$) individuals who have already had the disease and are therefore immune for life. 

Temporal changes in the numbers of individuals in different groups of the SIR model are described by the differential equations
\begin{subequations}
\begin{eqnarray}
\frac{dS}{dt}&=&-\beta SI, \label{eq1a}\\
\frac{dI}{dt}&=&\beta SI-aI, \label{eq1b}\\
\frac{dR}{dt}&=&aI \label{eq1c}.
\end{eqnarray}
\label{SIR}
\end{subequations}
The number of infected agents increases proportionally to the number of susceptible agents times the {\emph{force of infection}} $\beta I$, being itself the product of the non-zero infection rate $\beta$ and the number of infectious individuals. This is an example of a density dependent force of infection, being an alternative a frequency dependent force of infection $\beta I/N$~\cite{RockRPP14}. Individuals recover from the infection with rate $a$.

Adding Eqs.~(\ref{eq1a}-\ref{eq1c}), one can see that $S+I+R$ is time conserved. Furthermore, there is no notion of physical space in the model, meaning that individuals are uniformly mixed in the population. Both properties are common assumptions of simple SIR models that are rarely realized in real populations.

One of the key medical advances of the ${20}^{\text{th}}$ century has been the proliferation of cheap and safe vaccines for a range of diseases. Vaccines are among the most important tools to prevent infections and epidemics. One particular beneficial and mathematically interesting aspect of vaccination is that not $100\%$ of a population need to be immunized in order to prevent epidemic outbreaks, a property called herd immunity~\cite{RockRPP14}. Unfortunately, vaccination encounters opposition in different socio-cultural contexts that endangers the working of the herd immunity effect as the fraction of vaccinated individuals falls below a threshold. Yet another effect that limits the efficiency of vaccines is that the protection provided has a finite lifetime: depending on the disease, the immunization effect of the vaccine fades over time and the patient has to be re-vaccinated (e.g., diphtheria, tetanus and pertussis).

In this article, we investigate an  SIRV model (\enquote{V} for vaccination) that accounts for the fact that immunization of the vaccine wanes over time and that even recovered individuals can fall ill again. Furthermore, we consider the spatiotemporal dynamics of such a system. Epidemiological and SIRV models have been considered in many variants, for some reviews see Refs.~\cite{HethcoteSIAMR00,SchererBMB02,RockRPP14}, and some recent work consider the dynamics on networks~\cite{VerdascaJTB05,ShawPRE10,SunEPJB15}, information-driven vaccination~\cite{dOnofrioMMNP07,FengAAPSJ11,RuanPRE12}, or stochastic behavior~\cite{KamenevPRE08,IshikawaTISCIE12}. Spatiotemporal dynamics in nonlinear systems often show traveling wave patterns or Turing-like, stationary patterns~\cite{Walgraef97,MikSynI94}. In the context of spatial epidemiological models~\cite{RileyE15}, spatial coupling is often described by reaction-diffusion equations or networks~\cite{AbramsonBMB03,Rass03,NaetherEPJB08,LiuJSM07,HussainiJSM10,StancevicBMB13,CapalaEPJB17}.

\section{The SIRV model with finite lifetime protection}

\begin{figure}[!h]
\begin{center}
\includegraphics[width=0.75\textwidth]{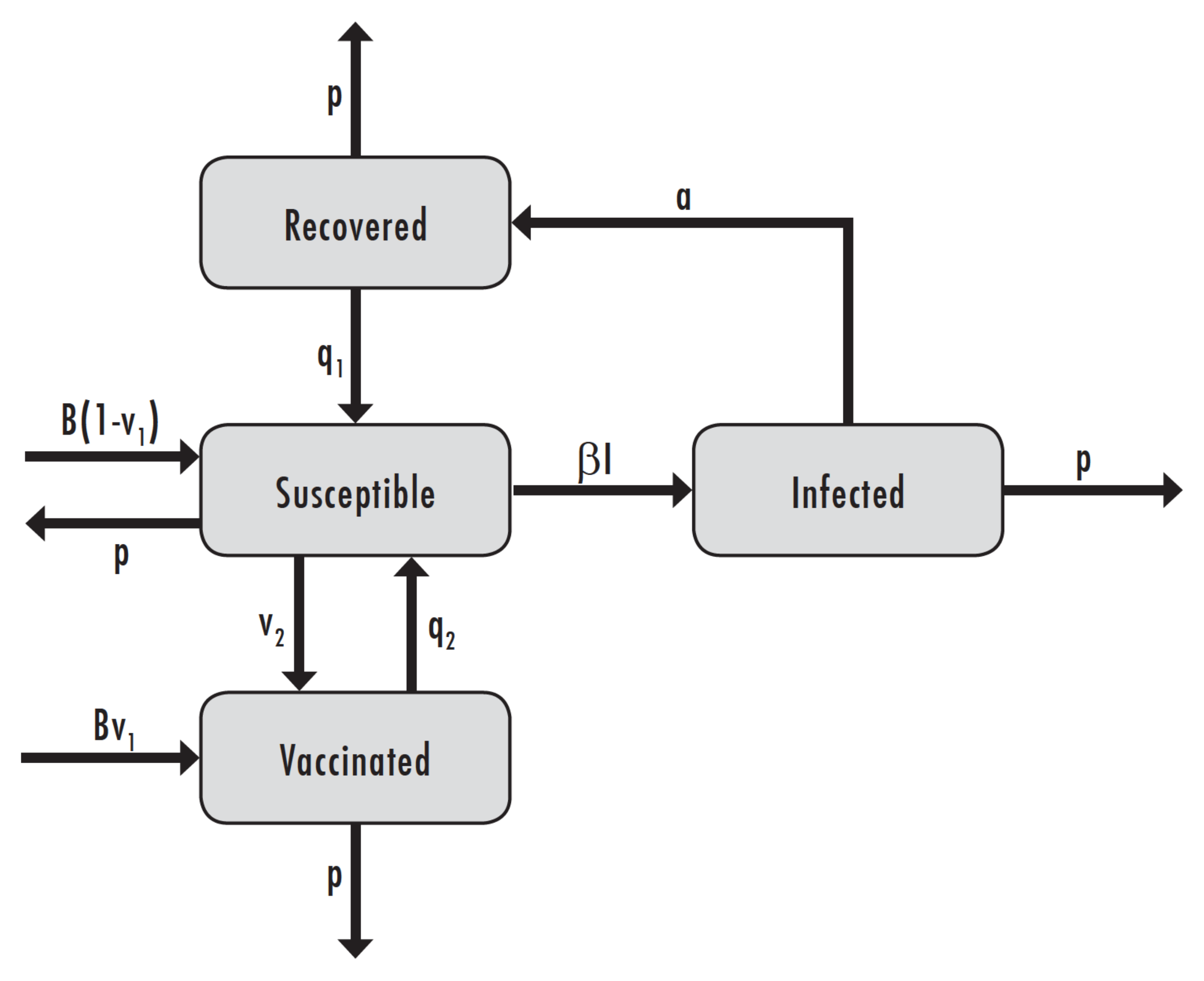}
\end{center}
\caption{Schematic representation of the transition of individuals in the SIRV model (SIR model with additional effect of vaccination) for infectious diseases.}
\label{fig:schemeSIRV}
\end{figure}

Among the range of epidemiological models using the SIR model as a building block, we focus on those aimed to investigate the impact of vaccination, using here a modification of the SIRV models presented in Refs.~\cite{SchererBMB02,YangJMAA10}. In this model, an independent birth rate $B$ and death rate $p$ are taken into account. There are two types of vaccination: $v_1$ is the fraction of newborns being vaccinated and $v_2$ is the rate of vaccination of susceptible individuals. By construction, $0\le v_1 \le 1$, with these two limiting cases representing that all newborns are either susceptible or vaccinated.  In contrary to the classical SIR model, protection against infection is not for life: recovered individuals become susceptible again with rate $q_1$, vaccinated ones with rate $q_2$, yielding the model
\begin{subequations}
\begin{eqnarray}
\frac{dS}{dt}&=&B(1-v_1)-\beta SI-v_2S + q_1 R + q_2 V -pS, \label{SIRVeq1}\\
\frac{dI}{dt}&=&\beta SI-aI-pI,\label{SIRVeq2}\\
\frac{dR}{dt}&=&aI-q_1 R-pR,\label{SIRVeq3}\\
\frac{dV}{dt}&=&v_1B+v_2S -q_2 V -pV\label{SIRVeq4}.
\end{eqnarray}
\label{SIRV}
\end{subequations}
A schematic representation of the transitions among the compartments in the SIRV model is displayed in Fig.~\ref{fig:schemeSIRV} and the meaning of the parameters is found in Table~\ref{table:parameters}. All parameters and variables have to be non-negative for Eqs.~(\ref{SIRV}) being interpreted as a valid epidemiological model.
%Fig 1 around here

\begin{table}
\begin{center}
\begin{tabular}{|l|c|}
\hline 
Parameter & Description\\
\hline
$\beta$ & Infection rate\\
\hline 
$a$ & Recovery rate from infection\\
\hline 
$p$ & Death rate\\
\hline 
$B$ & Birth rate\\
\hline 
$v_1$ & Fraction of vaccinated newborns\\
\hline 
$v_2$ & Vaccination rate of susceptible individuals \\
\hline 
$q_1$ & Conversion rate from recovered to susceptible \\
\hline
$q_2$ & Conversion rate from vaccinated to susceptible \\
\hline 
\end{tabular}
\end{center}
\caption{Parameters used in model~(\ref{SIRV}). The parameters $\beta$, $a$, $p$ and $B$ are positive, the others non-negative and $0\le v_1 \le 1$. All parameters are measured per unit time, except for $v_1$ (unitless) and $\beta$ (per unit time times population size unit).}
\label{table:parameters}
\end{table}

From Eqs.~(\ref{SIRVeq1}-\ref{SIRVeq4}), and defining the total population size as $N(t)=S+I+R+V$, we find
\begin{equation}
\frac{dN}{dt}=B-pN.
\end{equation}
This linear equation has the solution
\begin{equation}
N(t)=\frac{B}{p}+\left(N_0-\frac{B}{p}\right) \:e^{-pt},
\label{eq:Nt}
\end{equation}
where $N_0$ is the initial population size. While the total population size can vary as a transient if $N_0 \ne B/p $, in the limit $\displaystyle \lim_{t \to \infty}N= B/p$ which means that asymptotically the total population size is constant. In particular, if birth and death rates are equal, then $\displaystyle \lim_{t \to \infty}N= 1$. In the following, we require both birth and death rates to be strictly non-zero to exclude cases of vanishing or exponentially growing populations. Using Eq.~(\ref{eq:Nt}), the model in Eqs.~(\ref{SIRV}) can be interpreted as a time-dependent system with 3 variables (using, e.g., $R=N-S-I-V$). However, in the simulations, we numerically integrate the system defined in Eq.~(\ref{SIRV}) directly.

\section{Stationary states and their stability}

A key motivation of the study of population models is to find out the number of possible asymptotic states and to evaluate the relative size of the sub-populations. We find these steady state solutions in our model by setting the left-hand sides of Eqs.~(\ref{SIRV}) to zero. 

The first stationary state is characterized by the following solution for susceptible, infected, recovered and vaccinated agents:
\begin{eqnarray}
\{S_1,I_1,R_1,V_1\} &=& \left\{ \dfrac{B(q_2+p-p v_1)}{p (q_2+v_2+p)}, 0, 0, \dfrac{B(v_2+p v_1)}{p (q_2+v_2+p)}\right\}.
\label{eq:DFE} 
\end{eqnarray}
The number of infected individuals (and hence the number of recovered) is zero in this solution, justifying the name {\emph{disease-free equilibrium}} (denoted with subscript \enquote{1} and abbreviated as DFE). This solution describes a population without disease where the parameters control the relative fractions of susceptible and vaccinated individuals (summing up to $N=B/p$).

$S_1$ and $V_1$ are both independent of $q_1$, meaning that in the disease-free state the finite time of disease protection of a recovered individual is irrelevant (which is consistent with $R_1=0$). Neither the infection nor the recovery rate influences the steady state values $S_1$ and $V_1$. Note that $S_1$ cannot be negative since $0\le v_1\le 1$ (see above).  

The second stationary state is characterized by the following solution for susceptible, infected, recovered and vaccinated agents:
\begin{eqnarray}
\{S_2,I_2,R_2,V_2\} &=& \left\{  \dfrac{a+p}{\beta}, \dfrac{(q_1+p)[B\beta(q_2+p-pv_1)-p(a+p)(q_2+v_2+p)]}{\beta p(q_2+p)(a+q_1+p)},\right.\nonumber \\
&& \left. \dfrac{a}{q_1+p}I_2, \dfrac{B\beta v_1 +v_2(a+p)}{\beta(q_2+p)} \right\}.
\label{eq:ES} 
\end{eqnarray}
Since the number of infected individuals (and hence the number of recovered) is non-zero, this state (denoted with subscript \enquote{2}) is referred to as the {\emph{endemic equilibrium}} (EE). Again, only the parameters control the relative fractions of individuals in the different compartments (summing up to $N=B/p$). To describe a state relevant from a population dynamics point of view, the four compartments must have non-negative population numbers, constraining the parameter values, as we see later. 

%\section{Stability analysis of stationary states}

To perform a standard linear stability analysis, we determine the Jacobian of Eqs.~(\ref{SIRV}) and obtain
\begin{equation}
J=
  \begin{pmatrix}
    -(\beta I^{\ast}+v_2 +p) & -\beta S^{\ast} & q_1 & q_2\\
    \beta I^{\ast} & -(-\beta S^{\ast}+a+p) & 0 & 0\\
    0 & a & -(q_1+p) & 0\\
    v_2 & 0 & 0 & -(q_2 +p)
  \end{pmatrix},
\label{eq:Jacobian}
\end{equation}
where $I^{\ast}$ and $S^{\ast}$ need to be replaced by the respective stationary state solutions. Evaluated at the disease-free equilibrium, we obtain the eigenvalues as
\begin{subequations}
\begin{eqnarray}
\lambda_1 &=& -p,\label{eq:lambda1}\\
\lambda_2 &=& -q_1-p,\label{eq:lambda2}\\
\lambda_3 &=& -q_2-v_2-p,\label{eq:lambda3}\\
\lambda_4 &=& -a-p+ \dfrac{B\beta(q_2+p-pv_1)}{p(q_2+v_2+p)}=-a-p+ \beta S_1.\label{eq:lambda4}
\end{eqnarray}
\label{eq:eigenvalues}
\end{subequations}
%{\color{red}{MS: We can write $\lambda_4$ in at least 6 interesting ways:
%\begin{eqnarray}\lambda_4 &=& \dfrac{B\beta(q_2+p-pv_1)-p(a+p)(q_2+v_2+p)}{p(q_2+v_2+p)}\\
%&=& -a-p+ \dfrac{B\beta(q_2+p-pv_1)}{p(q_2+v_2+p)}\nonumber\\
%&=& -a-p+ \beta S_1 \nonumber\\
%&=&\dfrac{\beta(q_2+p)(a+q_1+p)}{q_2+v_2+p} I_2\nonumber\\
%&=& (a+p)(R_0-1)\nonumber\\
%&=& (a+p)\dfrac{\beta-\beta_c}{\beta_c}\nonumber\\
%\end{eqnarray}
%}}
The first three eigenvalues are always negative since all rates are non-negative and $p$ is strictly positive. The fourth eigenvalue can change sign (and therefore indicate instability of the solution), depending on the values of all parameters except $q_1$, that does not influence the stability of the disease-free state at the linear level. 

\begin{figure}[!h]
\begin{center}
\includegraphics[width=0.75\textwidth]{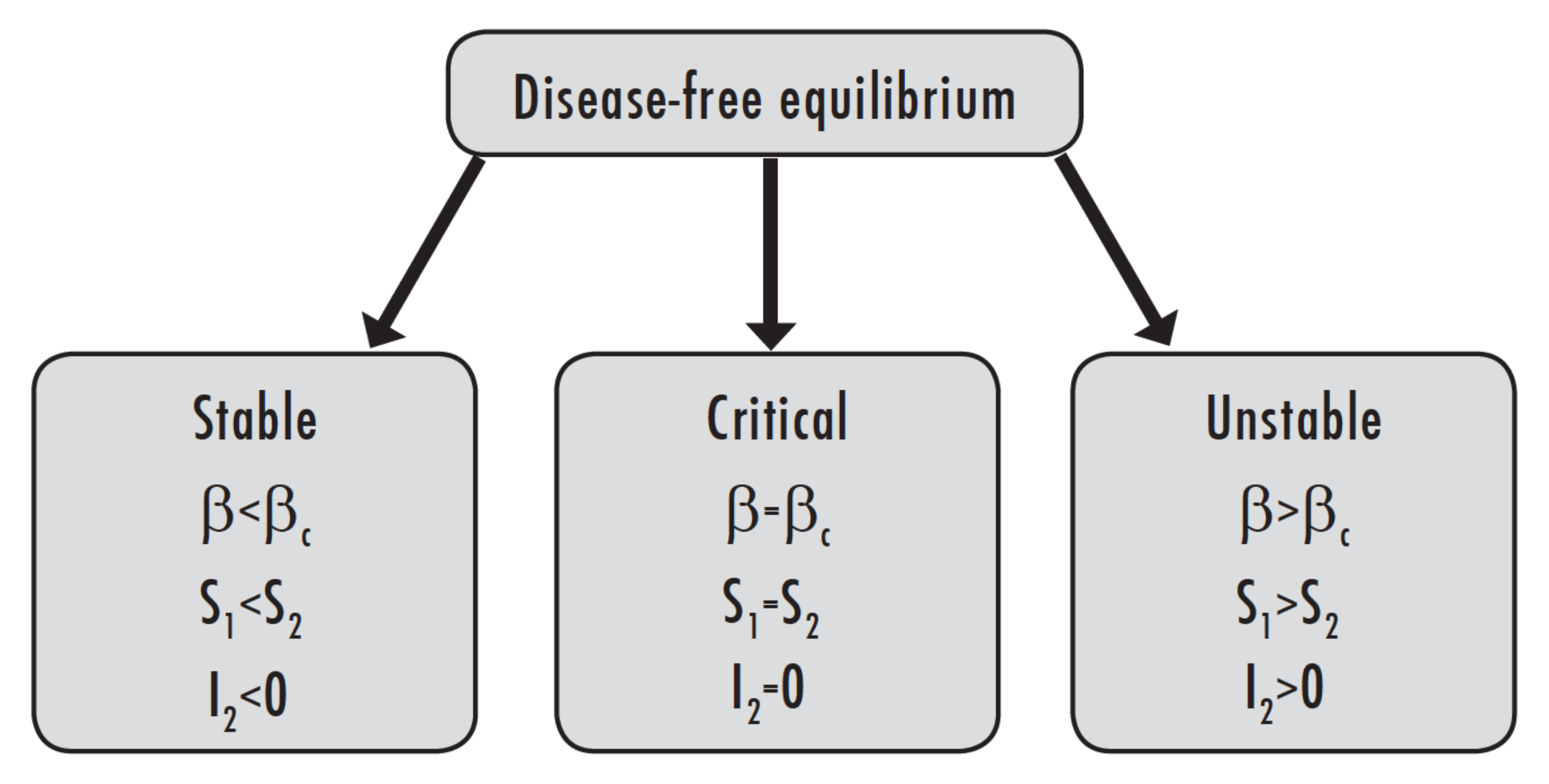}
\end{center}
\caption{Illustration of the stability of the disease-free state ($I_1=0$). Note that $I_2<0$ corresponds to a non-physical solution and hence absence of the endemic equilibrium state (population numbers have to be positive).}
\label{fig:stability}
\end{figure}

Setting $\lambda_4=0$, we obtain that the DFE is unstable if the infection rate $\beta$ is larger than a critical infection rate $\beta_c$, given by
\begin{equation}
%\beta_c=\frac{p}{B}\cdot\frac{p(q_2+v_2+a+p)+a(q_2+v_2)}{q_2+p-pv_1}, 
\beta_c=\frac{p(a+p)(q_2+v_2+p)}{B(q_2+p-pv_1)}.
\label{eq:betac}
\end{equation}
For $\beta=\beta_c$, one can show that $S_2= S_1$ while $I_2=0$. This is schematically shown in Fig.~\ref{fig:stability}.
Obviously, this condition can also be expressed as critical values for the other parameters (illustrated below).  The condition $\lambda_4=0$ coincides with the parameter value for which the DFE and the EE are identical. $\lambda_4$ can also be written as $\beta(q_2+p)(a+q_1+p)(q_2+v_2+p)^{-1} I_2$ which shows that the existence of the endemic equilibrium is associated with the instability of the disease-free equilibrium. The stability of the endemic equilibrium can be checked for in an analogous way but is omitted here as they lead to very lengthy expressions that have to be evaluated numerically. The existence of a stable DFE does not exclude the possibility of an appropriate initial condition mediated epidemic outbreak via a transient increase in the value of $I$.

%Fig 2 around here

\section{Transition between states and basic reproduction number}

In this section, we show how the stationary states vary as a function of some of the parameters. In particular, we consider the vaccination parameters $v_1$ and $v_2$ and the conversion rate $q_2$ (we have seen above that $q_1$ does not influence the existence or change of stability of the DFE).

\begin{figure}[!h]
\begin{center}
\includegraphics[width=0.98\textwidth]{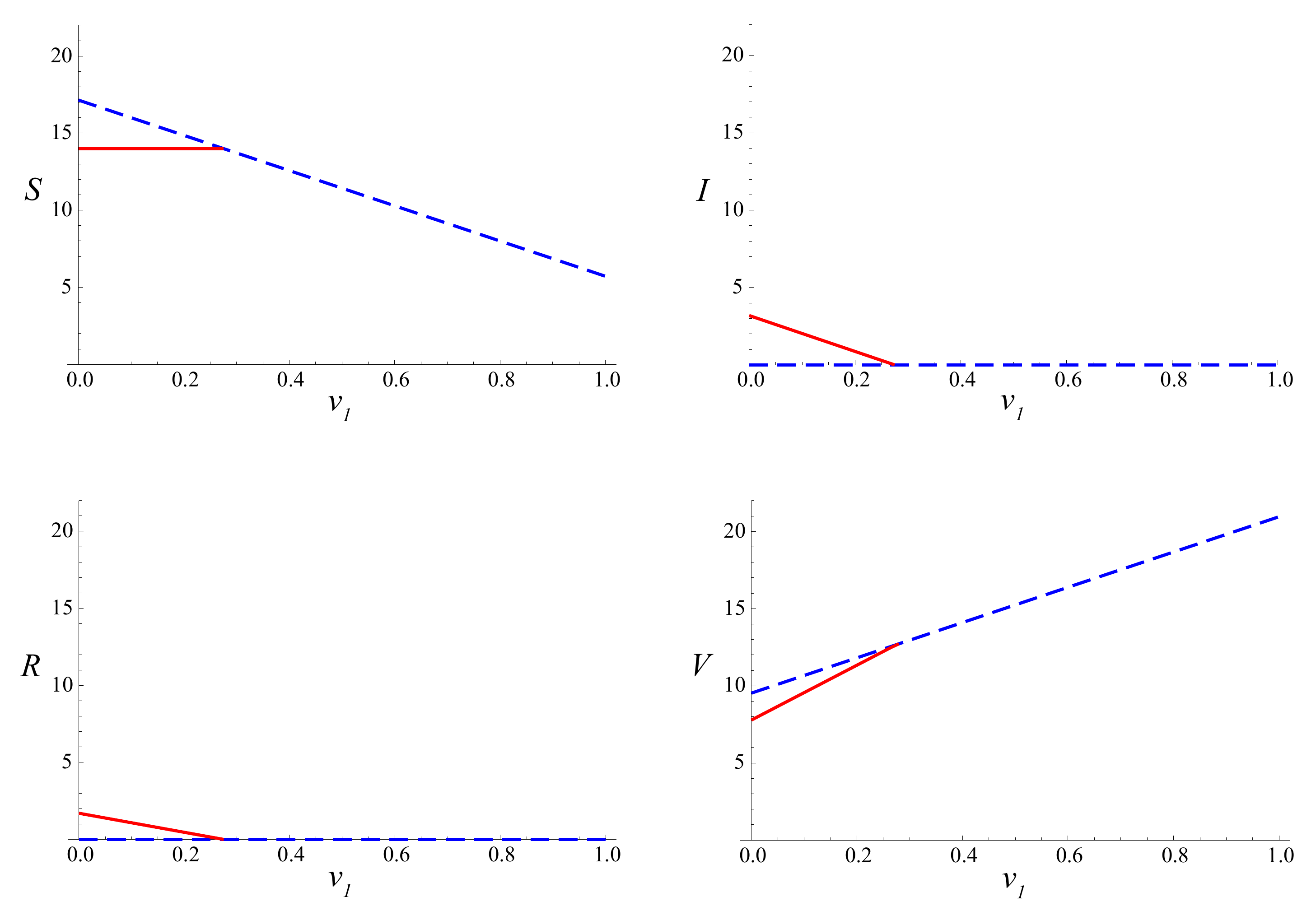}
\end{center}
\caption{The stationary states  as function of $v_1$: the red (solid) curve indicates the endemic state, the blue (dashed) curve the disease-free state. Other parameters used are as follows: $ \beta = 0.05$, $a = 0.4$, $p = 0.3$, $B = 8.0$, $v_2 = 0.25$, $q_1 = 0.45$, $q_2 = 0.15$.}
\label{fig:bifv1}
\end{figure}

Figure~\ref{fig:bifv1} shows the stationary state solutions for all four sub-populations as a function of the fraction of vaccinated newborns ($v_1$). For the DFE, the dependence on $v_1$ is linear for $S_1$ and $V_1$, showing the direct proportionality of the fraction of vaccinated people in the population on the fraction of vaccinated newborns. As $v_1$ is decreased. the DFE becomes unstable at a critical $v_{1c}$ via a transcritical bifurcation and the EE sets in, a general feature of SIR models~\cite{HethcoteSIAMR00}. Then, the number of susceptible remains constant in the population while the number of infected (and also recovered) increases linearly. At the same time, the vaccinated fraction of the population decreases, and with a higher rate than when the DFE was stable.
%Fig 3 around here

\begin{figure}[!h]
\begin{center}
\includegraphics[width=0.98\textwidth]{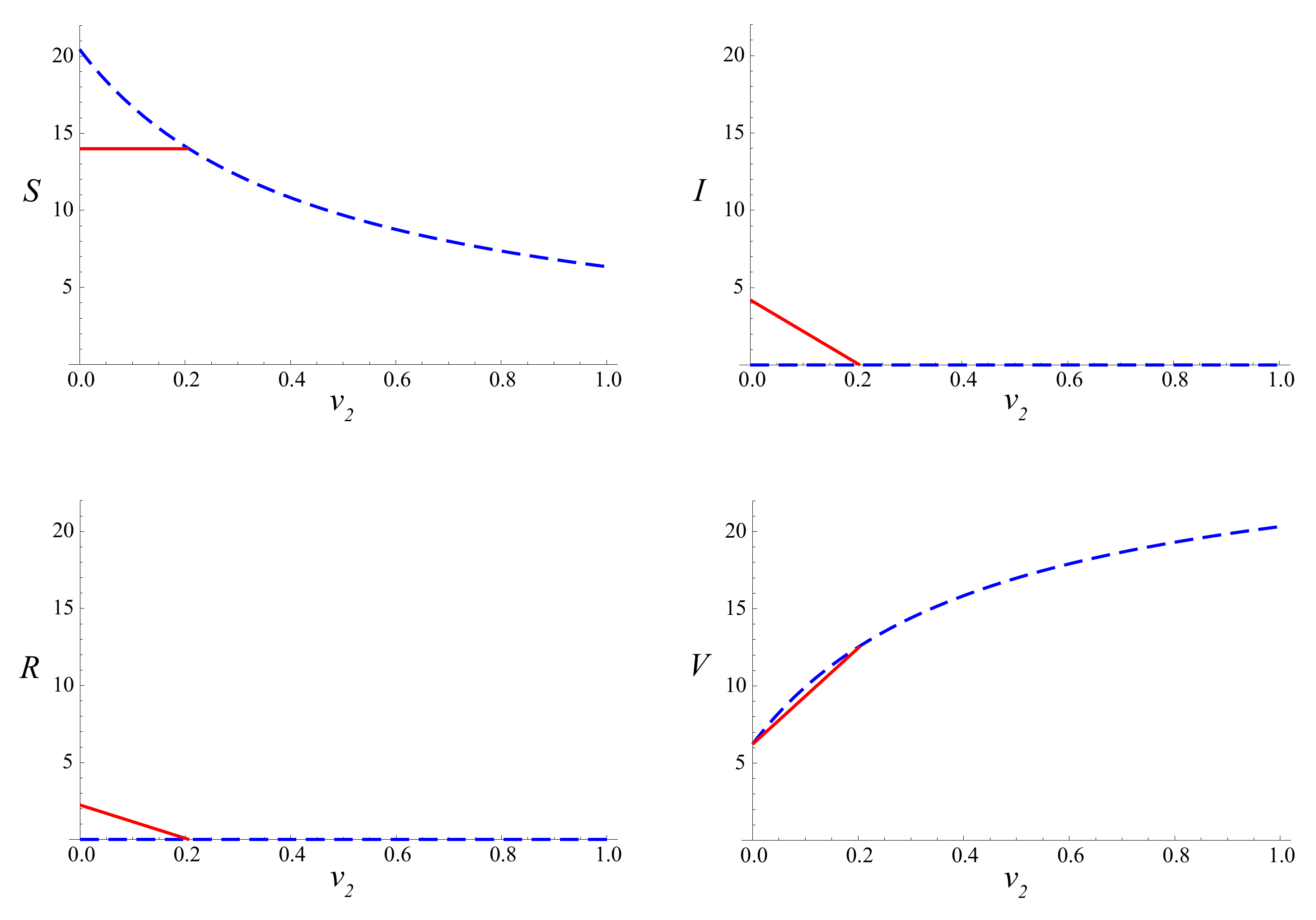}
\end{center}
\caption{The stationary states as function of $v_2$: the red (solid) curve indicates the endemic state, the blue (dashed) curve the disease-free state. Other parameters used are as follows: $\beta = 0.05$, $a = 0.4$, $p = 0.3$, $B = 8.0$, $v_1 = 0.35$, $q_1 = 0.45$, $q_2 = 0.15$.}
\label{fig:bifv2}
\end{figure}

We now consider the case of varying the vaccination rate of the susceptible individuals $v_2$ (Fig.~\ref{fig:bifv2}). Considering first the EE, it can be seen that the qualitative behavior of the curves is similar to the case of varying the fraction of vaccinated newborns. However, for the DFE the fractions of susceptible and vaccinated sub-populations do not change linearly as above, see also Eq.~(\ref{eq:DFE}). In particular, the rate of increase of $V_1$ as a function of $v_2$ starts slowing down beyond the linear regime, meaning that it becomes increasingly difficult to protect the population. Also, for $v_2=0$, $V_1=V_2$ and hence if the only vaccination is taking place at birth, the fraction of vaccinated people is identical in the endemic and disease-free states.
%Fig 4 around here

Finally, we discuss the case of changing the conversion rate from vaccinated to susceptibles ($q_2$). The loss of protection of the vaccination plays an antagonistic role to the vaccination rate. It is not a surprise to find that for the DFE, the vaccinated fraction of the population decreases as $q_2$ increases. The role of $q_2$ in the equation for $S_1$ is the same as $v_2$ in the equation for $V_1$ [Eq.~(\ref{eq:DFE})]. For the EE, though, the situation is slightly different. While the infected fraction of the population increases with $q_2$, it does so at rate that is slower than the linear growth rate at onset, showing that a waning immunization favors the endemic state, but a change in this parameter is less dangerous than a decrease of any of the vaccination parameters.
%Fig 5 around here

A relevant quantity in epidemiology is the {\emph{basic reproduction number}} $R_0$. It is defined as the expected number of secondary individuals infected by an individual in its lifetime (for a review see Ref.~\cite{HeffernanJRSI05}). By analyzing this value it is possible to predict whether a disease present in a population will create an epidemic (if $R_0>1$). 

To calculate the basic reproduction number $R_0$, we use the next generation method for structured populations~\cite{HeffernanJRSI05}. For that we separate the Jacobian given in Eq.~(\ref{eq:Jacobian}) into a transmission part $T$ and transition part $\Sigma$, evaluated at the DFE. We obtain:
\begin{equation}
T=
  \begin{pmatrix}
    0 & -\beta S_1 & 0 & 0\\
    0 & \beta S_1  & 0 & 0\\
    0 & 0 & 0 & 0\\
    0 & 0 & 0 & 0
  \end{pmatrix}
\label{eq:T}
\end{equation}
and
\begin{equation}
\Sigma=
  \begin{pmatrix}
    -v_2 -p & 0 & q_1 & q_2\\
    0 & -a-p & 0 & 0\\
    0 & a & -q_1-p & 0\\
    v_2 & 0 & 0 & -q_2 -p
  \end{pmatrix}.
\label{eq:Sigma}
\end{equation}
Then, $R_0$ is the leading eigenvalue of the matrix $[-T\Sigma^{-1}]$. It is determined as
\begin{equation}
R_0=\frac{B\beta(q_2+p-pv_1)}{p(a+p)(q_2+v_2+p)}.
\label{eq:R0}
\end{equation}
The $R_0$ shown in Eq.~(\ref{eq:R0}) above is identical to $S_1/S_2$ and to $\beta/\beta_c$, providing alternative interpretations of the onset of an epidemic. Also, $R_0=1+(a+p)^{-1}\lambda_4$, elucidating the relationship between the basic reproduction number and the dominant eigenvalue of the stability analysis of the DFE. Because of this link it is not surprising that for this model, the same result can be obtained by evaluating $\lambda_4$ or by setting $I_2$ to zero.

Figure~\ref{fig:R0} shows $R_0$ as a function of the vaccination parameters $v_1$, $v_2$, and the rate of loss of protection $q_2$. The panels A-C exhibit a situation involving an epidemic with low $R_0$, while panels D-F use parameter values for pertussis, a disease with high $R_0$. In agreement with the above figures for the stationary states, we observe that for low vaccination rates ($v_1,\:v_2$) and a high rate of loss of protection ($q_2$), the endemic state is stable while the disease-free state is unstable. On the other hand, if the vaccinated fraction of the population fast loses its protection, a transition from the disease-free state to the endemic equilibrium occurs. Only the dependence of $R_0$ on $v_1$ is linear. As the infection rate $\beta$ increases, the $R_0$ curves are shifted upwards (panels A-C), reflecting an increased tendency to destabilize the disease-free equilibrium. For the specific case of pertussis, we observe that the curve $R_0(v_1)$ is relatively flat (for three different combinations of values of $v_2$ and $q_2$), meaning that even complete vaccination of a newborn population is not sufficient to contain the disease if the vaccination rate $v_2$ is not high enough. This is confirmed by the curve $R_0(v_2)$ (Fig.~\ref{fig:R0}E) which shows a sharp decay, illustrating that vaccination is an efficient way of decreasing $R_0$ (for three different combinations of values of $v_1$ and $q_2$). It is worth noting that red and blue curves differ only in $v_1$ and while the values are very different (0.95 and 0.05), the position of the curves are similar. In contrast to this, increasing $q_2$ from 0.05 to 0.3 make the control via $v_2$ more difficult (purple curve). Finally, the curves in Fig.~\ref{fig:R0}F show that $R_0$ is also very sensitive to the conversion rate from vaccinated to susceptible as only small values allow that  $R_0$ stays below 1.

%Fig 6 around here

\begin{figure}[!h]
\begin{center}
\includegraphics[width=0.98\textwidth]{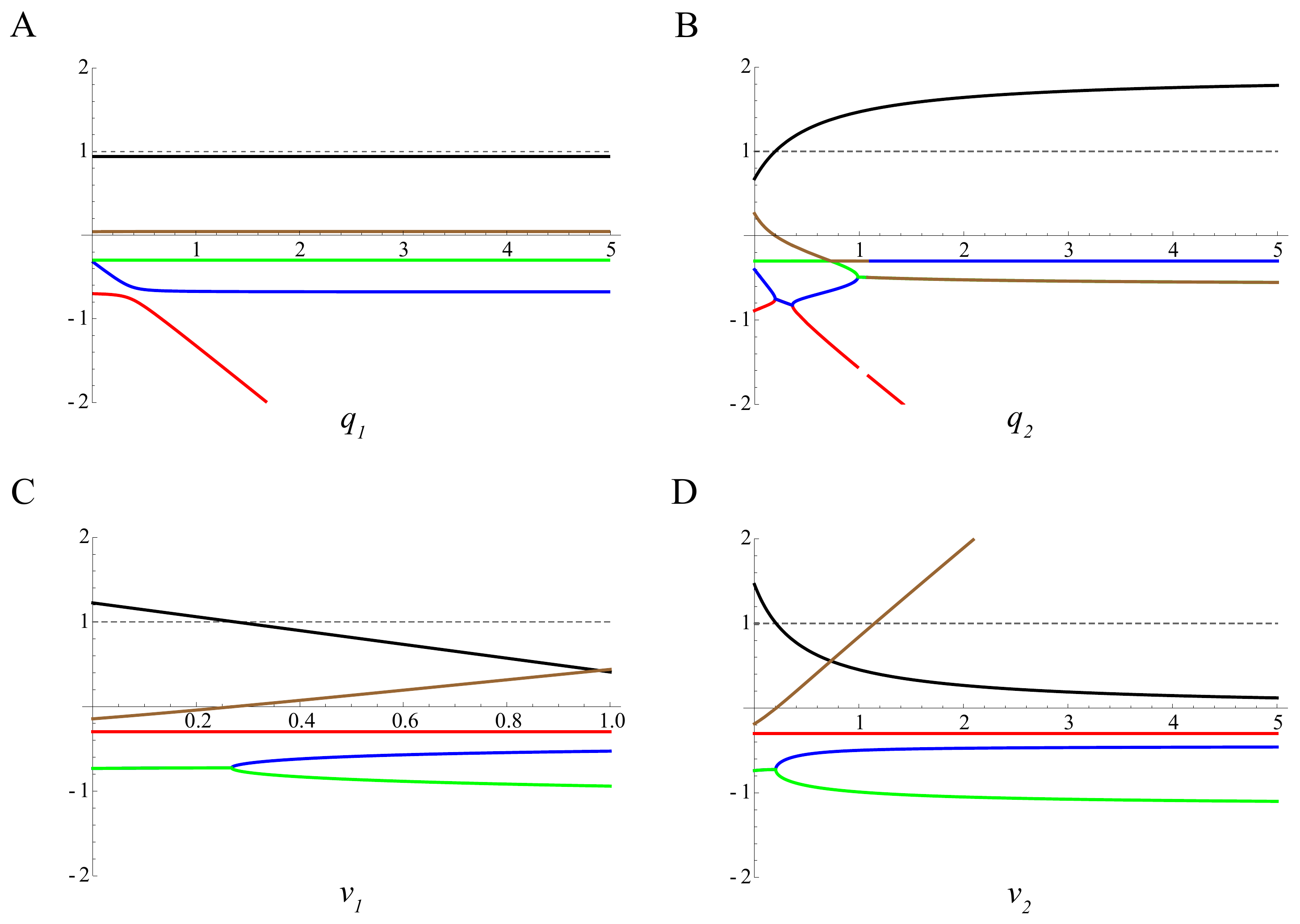}
\end{center}
\caption{Stability of the endemic equilibrium. We show the real parts of the four eigenvalues as function of the parameters $q_1$ (A), $q_2$ (B), $v_1$ (C) and $v_2$ (D), together with the curve for $R_0$ (black) indicating the stability of the DFE (the dotted line at $R_0=1$ is a guide to the eye). Where the DFE is unstable, the EE is stable. Other parameter values are as follows: $a = 0.4$, $\beta = 0.05$, $p = 0.3$, $B = 8.0$, $v_1 = 0.35$, $v_2 = 0.25$, $q_1 = 0.45$, $q_2= 0.15$.}
\label{fig:EEstab}
\end{figure}

As mentioned above, the stability analysis of the endemic equilibrium leads to very lengthy expressions that we exclude for the sake of brevity. We, hence, assess the stability of the endemic equilibirium numerically. In Fig.~\ref{fig:EEstab} we show how the four eigenvalues vary as a function of the main parameters $q_1$, $q_2$, $v_1$ and $v_2$. As we know that the EE only exists if the DFE is unstable, we also plot the $R_0$ curve indicating the critical parameter values. The fundamental result is that where $R_0>1$, the real parts of the four eigenvalues are negative, showing that the EE is stable in these parameter regions. A particular case is the graph Fig.~\ref{fig:EEstab}A which confirms that $q_1$ is not only irrelevant for the stability of the DFE, but also of the EE.
%Fig 7 around here

While the existence and stability of the asymptotic states are fundamental properties of any epidemiological system, an epidemic is a time-dependent process. For example, even if the DFE is stable and asymptotically obtained, an epidemic outburst can occur. Hence, in a deterministic system, the initial state of the population is fundamentally important. In the presence of birth-death processes, a high death rate can mask slow processes (if the loss of immunization is on the timescale of life expectancy) or an expanding population may require a higher vaccination rate in order to keep the population protected.

\section{The spatiotemporal SIRV model}

Epidemiological models without spatial degrees of freedom can only be applied to very well mixed populations. However, people live in confined communities that are spatially connected. As a starting approach, we assume that the population is distributed over a one-dimensional space where transport between adjacent areas is diffusive (equivalent to nearest-neighbor interactions). Therefore, we add diffusion terms to the SIRV model~(\ref{SIRV}) and obtain
\begin{subequations}
\begin{eqnarray}
\frac{\partial S}{\partial t}&=&B(1-v_1)-\beta SI-v_2S + q_1 R + q_2 V -pS+D_S\nabla^2S, \label{SIRVDeq1}\\
\frac{\partial I}{\partial t}&=&\beta SI-aI-pI+D_I\nabla^2I,\label{SIRVDeq2}\\
\frac{\partial R}{\partial t}&=&aI-q_1 R-pR+D_R\nabla^2R,\label{SIRVDeq3}\\
\frac{\partial V}{\partial t}&=&v_1B+v_2S -q_2 V -pV+D_V\nabla^2V\label{SIRVDeq4},
\end{eqnarray}
\label{eq:SIRVD}
\end{subequations}
where $ D_F$ ($F=S,I,R,V$) are diffusion constants for susceptible, infected, recovered and vaccinated individuals, respectively.

The two fixed points shown in Eqs.~(\ref{eq:DFE}) and~(\ref{eq:ES}) of the diffusion-free system are solutions of the system with diffusion~(\ref{eq:SIRVD}) in case the variables do not show any spatial dependence, i.e., represent a {\emph{homogeneous}} solution. However, the linear stability of this homogeneous solution depends on diffusion, as we shall see now.

Perturbed around the homogeneous fixed points, in the Fourier transformed $({\bf k},t)$ space, the dynamics is represented through the Jacobian $J_k$ that is given by
\begin{equation}J_k=\left( \begin{array}{cccc} 
-(\beta I^{\ast} + v_2 + p + D_S k^2) & -\beta S^{\ast} & q_1 & q_2 \\
\beta I^{\ast} & -(-\beta S^{\ast}+a +p +D_I k^2) & 0 & 0 \\
0 & a & -(q_1+p+D_R k^2) & 0 \\
v_2 & 0 & 0 & -(q_2+p+D_V k^2) \\
\end{array} \right),
\end{equation}
where $I^{\ast}$ and $S^{\ast}$ need to be replaced by the respective stationary state solution [Eqs.~(\ref{eq:DFE},~\ref{eq:ES})]. Around the disease-free steady state, one can find the eigenvalues as follows
\begin{subequations}
\begin{eqnarray}
\lambda_1(k) &=& -q_1-p-D_R k^2, \label{eq:lambda1k}  \\
\lambda_2(k) &=& -a-p+ \dfrac{B\beta(q_2+p-pv_1)}{p(q_2+v_2+p)}-D_I k^2, \label{eq:lambda2k}  \\
\lambda_3(k) &=& \frac{1}{2} \left[ -2p-(D_S + D_V) k^2 -q_2-v_2+ \right.\nonumber \\ & & +\left. \sqrt{{[q_2-(D_S-D_V) k^2]}^2 +2[(D_S-D_V)k^2+q_2]v_2+{v_2}^2} \right],\label{eq:lambda3k} \\
\lambda_4(k) &=&  \frac{1}{2} \left[ -2p-(D_s + D_V) k^2 -q_2-v_2-\right.\nonumber \\ & & -\left.\sqrt{{[q_2-(D_S-D_V) k^2]}^2 +2[(D_S-D_V)k^2+q_2]v_2+{v_2}^2} \right].
\label{eq:lambda4k}
\end{eqnarray}
\label{eq:eigenvaluesk}
\end{subequations}
The eigenvalue $\lambda_1(k)$ is a generalization of $\lambda_2$ of the ODE system and is always negative. The eigenvalue $\lambda_2(k)$ is the generalization of $\lambda_4$ of the ODE system and can therefore change sign. The eigenvalues $\lambda_{3,4}(k)$ depend on the sum and differences of the diffusion coefficients of the susceptible and vaccinated population fractions. It can be easily shown that $\lambda_{3,4}(k)$ are always negative and hence diffusion has always a stabilising effect on the DFE. The most unstable wavenumber is $k=0$. Hence, whenever the DFE is unstable in the purely temporal system, it is also unstable in the spatiotemporal system.

\begin{figure}[!h]
\begin{center}
\includegraphics[width=0.98\textwidth]{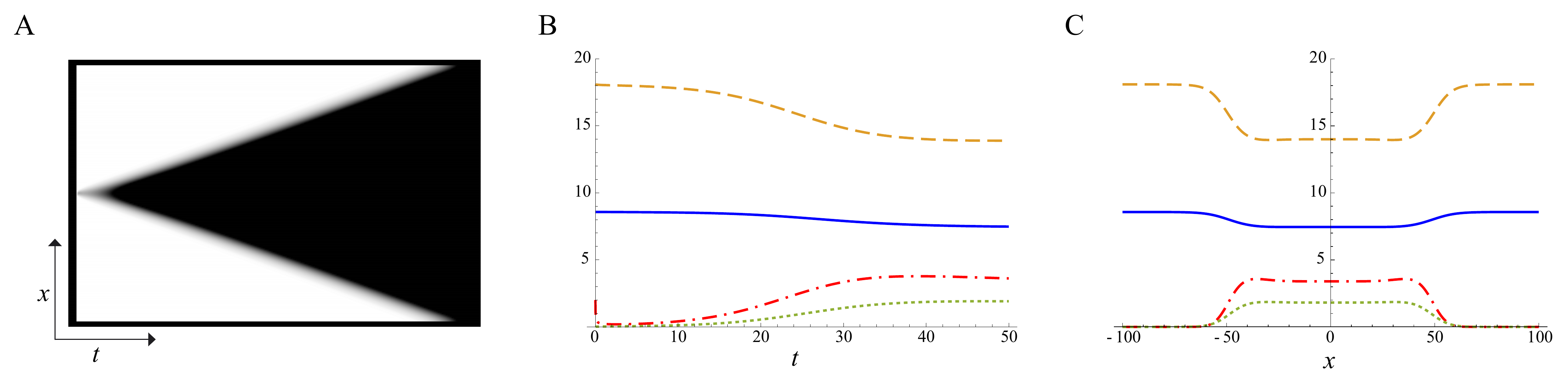}
\end{center}
\caption{Wave of an epidemic spread as observed in the SIRV model. (A) Space-time density plot for $I$. (B) Temporal variation of $\{S, I, R, V\}$ at the centre of the simulation domain considered ($x=0$). (C) Front profile of $\{S, I, R, V\}$ at $t=100$. The brown (dashed) curve denotes $S$, the red (dot-dashed) curve denotes $I$, the green (dotted) curve denotes $R$ and the blue (solid) curve represents $V$. Parameters used are as follows: $ \beta = 0.05$, $a = 0.4$, $p = 0.3$, $B = 8.0$, $v_1 = 0.25$, $v_2 = 0.15$, $q_1 = 0.45$, $q_2= 0.25$, $D_S=10$, $D_I=0.5$, $D_R=10$, $D_V=10$. The system size is $-100\le x\le 100$, the boundary conditions are periodic and the displayed time interval in (A) is $T=180$.}
\label{fig:simspatial}
\end{figure}

Figure~\ref{fig:simspatial} shows a numerical simulation of the spatiotemporal SIRV model [Eqs.~(\ref{eq:SIRVD})] in one-dimensional space. The initial condition is a disease-free state with a small nucleus of infected agents at the center of the medium. Parameters have been chosen to ensure that the disease-free state is unstable, leading to a transition to the endemic state. This can be clearly seen as a traveling wave in the space-time plot for $I$ in Fig.~\ref{fig:simspatial}A. Figures~\ref{fig:simspatial}B and~\ref{fig:simspatial}C illustrate the behavior of all variables for a fixed point in space (B) and for a fixed point in time (C). The latter displays the profile of the traveling wave front. For this set of parameters, the spatial distribution for $I$ shows small peaks in the fronts.
%Fig 8 around here

\begin{figure}[!h]
\begin{center}
\includegraphics[width=0.7\textwidth]{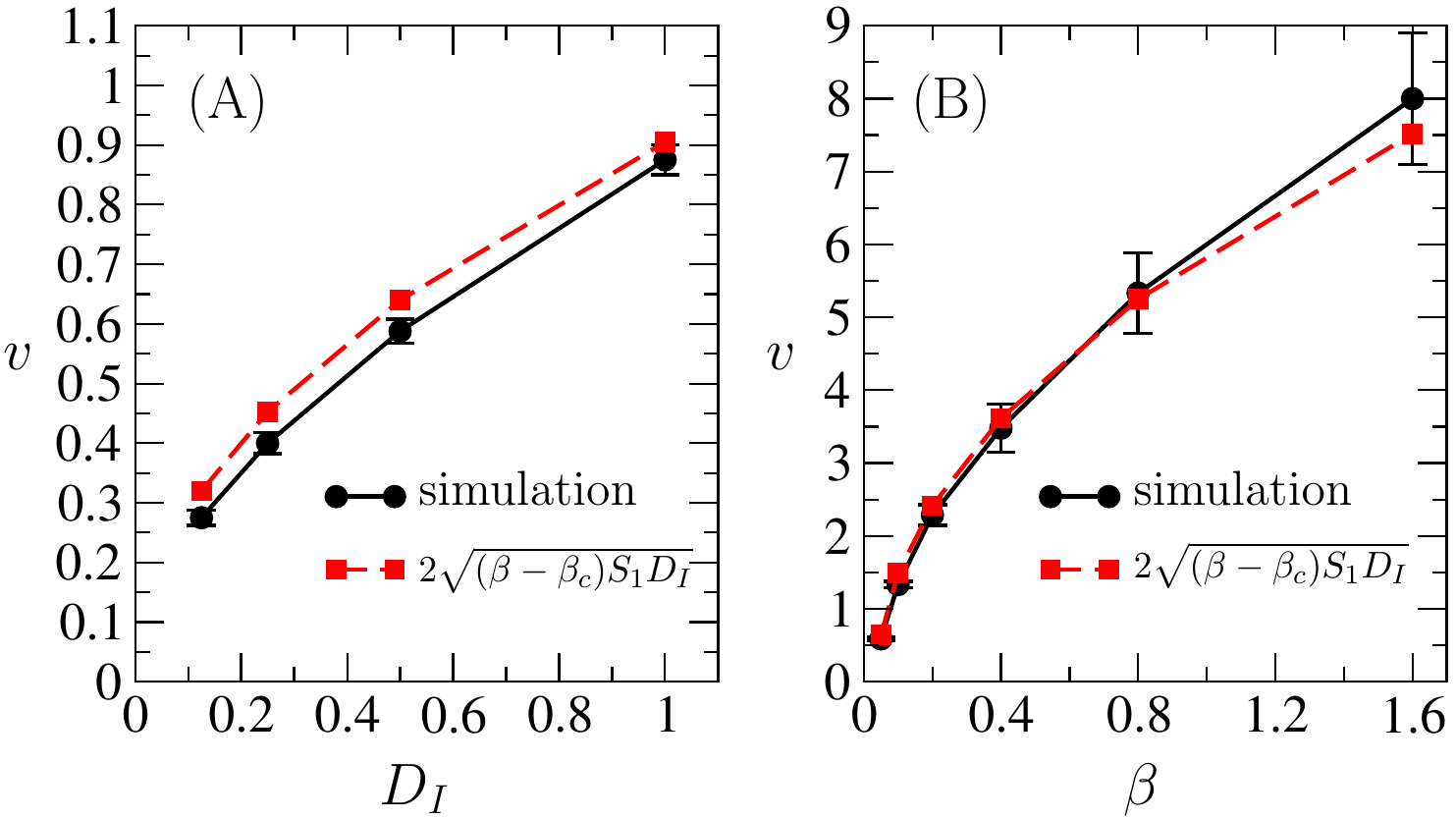}
\end{center}
\caption{Wave speed of an epidemic in the SIRV model as a function of $D_I$ (A) and $\beta$ (B). Speeds are measured from the simulation data (black solid curves) and compared with Eq.~(\ref{eq:vmin}) (red dashed curves). No fitting parameters are used. Other parameters are as in Fig.~\ref{fig:simspatial}.}
\label{fig:vmin}
\end{figure}

The wave of infection observed in Fig.~\ref{fig:simspatial} can be investigated in more detail. In Fig.~\ref{fig:vmin}, we show how the front velocity changes with the diffusion constant $D_I$ and the infection rate $\beta$. Both functional forms follow a square root dependence reminiscent of the Fisher-Kolmogorov equation~\cite{MikSynI94,Murray89}. Indeed, for a single-species population model with variable $u$, it is known that the natural front velocity of a front triggering a transition from the unstable to the stable state is given by $v=2\sqrt{ f'(u_1)D}$, where $D$ is the diffusion constant, $f(u)$ describes the temporal dynamics and $u_1$ is the unstable steady state~\cite{Murray89}. Applying the same rationale to Eq.~(\ref{SIRVeq2}), we obtain 
\begin{equation}
v=2\sqrt{(\beta S_1-a-p)D_I}=2\sqrt{(\beta-\beta_c)S_1D_I}.
\label{eq:vmin}
\end{equation}
The qualitative agreement between the curves is surprisingly good which is remarkable as no fitting parameters are applied and the analytic expression uses only one equation of a coupled 4-dimensional dynamical system. There is a slight quantitative difference for small velocities, as seen in Fig.~\ref{fig:vmin}A that could be partially explained by the fact that the simulations are performed in a finite sized system and that the calculation of the front speed from the simulation data carries an error.
%Fig 9 around here

\section{Discussion}

In this article, we have considered an SIRV model in the temporal and spatiotemporal domain. The model has two asymptotic states, the disease-free state and the endemic state. We have focused on the consequences of diminishing immunization, i.e., the effect when vaccinated or recovered individuals become susceptible again. The results have been obtained through bifurcation analysis of the individual solutions (for $S$, $I$, $R$ and $V$), as well as through the determination of the basic reproduction number $R_0$. In the asymptotic regime the number of each sub-populations is proportional to its density in the whole population, so the results refer directly to population densities or fractions. Our exclusively temporal model shares similarities with a model studied in~\cite{YangJMAA10}, however, the models only coincide if we set $v_1=q_1=0$ in our model and simultaneously set $\mu=\sigma=0$ in the model discussed in Ref.~\cite{YangJMAA10}. However, assuming non-zero values for these parameter is crucial for both our model (possible vaccination at birth and conversion from recovered to susceptible agents) and the model discussed in Ref.~\cite{YangJMAA10} (variable vaccine efficacy and possibility of disease-induced deaths) and hence the interpretation and applicability of the models differ substantially.

By considering the results of a linear stability analysis of the disease-free state, we have found that the loss of protection of the {\emph{recovered}} fraction of the population (with rate $q_1$) has no influence on the onset of the endemic state. While the rate $q_1$ does not influence the asymptotic DFE, it can impact on the transient time to equilibrium. On the other hand, the loss of protection of the {\emph{vaccinated}} fraction of the population (with rate $q_2$) {\emph{can}} shift the population from a disease-free state to an endemic one.  An interesting feature of this model is that the density of susceptibles in the endemic regime does not depend on $q_2$. The curve of $R_0$ with $q_2$ is increasing, however, with a decreasing slope, meaning that decreasing $q_2$ in the epidemic regime may bring the population closer to the threshold than predicted by a linear regression. Considering the effect of the vaccination rates, we find that the fraction of vaccinated newborns $v_1$ changes the asymptotic fractions linearly in both stationary states, as well as $R_0$. This is in contrast to $v_2$ where the dependence is nonlinear. There, we have found that if $v_2$ is decreased in an disease-free state, the basic reproduction number increases more strongly than predicted by a linear regression. This implies that the critical $R_0=1$ may be reached for higher $v_2$ than assumed. 

Our results show that in the diffusion-free state the dominant eigenvalue of the disease-free state $\lambda_4=\dfrac{\beta(q_2+p)(a+q_1+p)}{q_2+v_2+p}\: I_2$. This means that if the endemic state exists, the disease-free equilibrium is unstable that is associated with positive $\lambda_4$ values and $R_0>1$. In the complete absence of adult vaccination, implying $v_2 =0$, Eqs.~(\ref{eq:DFE}) and~(\ref{eq:ES}) show that the vaccinated number density is the same for the two asymptotic states $\left(\dfrac{B v_1}{q_2 +p} \right)$. This then implies that one cannot predict the actual epidemic state from the proportion of the vaccinated population alone. We have presented a numerical solution for the stability problem of the endemic equilibrium. It indicates that while the EE exists, it is stable.

\begin{figure}[!h]
\begin{center}
\includegraphics[width=0.7\textwidth]{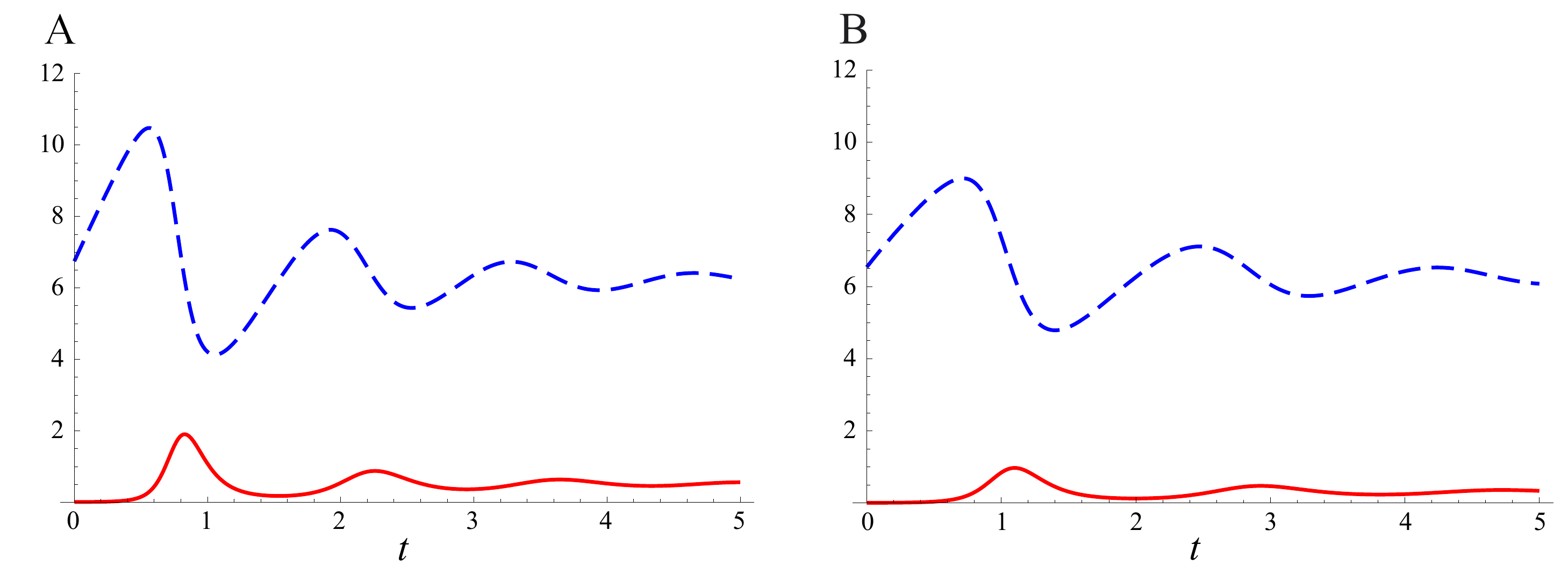}
\end{center}
\caption{Initial temporal dynamics of the fractions of infected (solid curve) and susceptibles (dashed curve) for a parameter set for pertussis (time unit is years): $p=1/70$, $r=140$, $a=365/23$, $B=0.026$, $v_1=0.95$, $q_1=0.1$, $q_2=0.05$ and $v_2=0$ (A) and $v_2=0.3$ (B). The dynamics is characterised by damped oscillations before settling to the endemic equilibrium.}
\label{fig:osc}
\end{figure}

The features obtained from a study of this model can be put in the appropriate context of epidemiological data. Diphtheria and Pertussis (whooping cough) are amongst the diseases that are associated with waning immunization. Repeated vaccinations (\enquote{boosts}) are needed to prevent the spreading of such diseases. Due to the high $R_0$ values of these diseases, children are vaccinated at early ages.  Without epidemiological control, the $R_0$ of pertussis has been estimated at 16-18~\cite{Anderson91}, a value lowered to 11-15~\cite{WearingPLoSP09} later. In the presence of vaccination, the value could be lowered to around 5.5~\cite{KretzschmarPLoSMED10}. The incidence among adults are explained by waning immunization and the possibility of evolving subclinical strains that are held responsible for persistence of pertussis in vaccinated populations~\cite{KretzschmarPLoSMED10}. In Fig.~\ref{fig:osc}, we show a short time series of a population suffering from pertussis infection and for which the endemic equilibrium is stable. The initial state consists of a population with very few infected agents. We clearly notice some outbursts of infection, with a characteristic time gap between 1 and 2 years. This timescale is not far off from known deterministic models of pertussis which consistently predict annual epidemics~\cite{NguyenJRSI08}. Note, however, that detailed and more realistic models for pertussis rely on an SEIR mechanism, with an exposed/latent phase and/or age structure, and possibly term-time forcing. Furthermore, stochastic effects are also known to be crucial in the disease dynamics~\cite{RohaniAN02}. A recent work compares the different classes of models including reinfection of recovered and loss of infection-derived immunity and subsequent reinfection~\cite{RozhnovaJRSI12}. In the context of this article, we simply want to illustrate an example of a specific disease for our model.

%Fig 10 around here

For all realistic epidemiological models, spatial interactions have to be considered. In our model, we have assumed a nearest-neighbor interaction, modeled by diffusion terms. Based on the definition of a spatial basic reproduction number $R_{0k}$, our linear stability analysis of the DFE confirms that the most unstable wavenumber is $k=0$, and that the disease-free equilibrium cannot be destablized by controlling the diffusion rates. A discussion of the spatial stability problem of the endemic equilibrium is beyond the scope of the present work.

A well-known feature of infection models with diffusion is that they are able to describe the propagation of waves, of particular interest being waves that represent the onset of an epidemic. We have shown that in spite of the comparatively high complexity of the model (4 coupled equations), the wave speed still approximately follows the one-species Fisher-Kolmogorov model, similar to what has been observed for a different model~\cite{AbramsonBMB03}.

Temporal and spatiotemporal epidemiological models have been studied in many variants. A series of recent works try to find optimal vaccination strategies, for example by a probabilistic modelling of infection in networks~\cite{TakeuchiJTB06}, by minimizing the number of infected and susceptibles~\cite{ZamanBioS08}, by a Poisson distributed vaccination schedule on networks~\cite{ShawPRE10}, by an information (and time) dependent vaccination rate~\cite{RuanPRE12}, or by optimizing the vaccination rate through a stochastic maximum principle~\cite{IshikawaTISCIE12}. In contrast to these articles, we analyze the front speed of a general SIRV model, similar to the approaches of~\cite{NaetherEPJB08} for a stochastic SIR model and~\cite{HussainiJSM10} for an SIR model with non-smooth treatment (vaccination) functions.

As an outlook to further work in the model, we mention the spatial stability analysis of the endemic equilibrium, an analytical and numerical investigation of fronts in two space dimensions and the incorporation of social effects.

%\acknowledgments{}

\section*{Acknowledgments}
%\begin{acknowledgements}
The authors acknowledge some preliminary simulations by Asad Ahmad. M.S. would like to thank EPSRC for funding via the grant EP/M02735X/1.

All authors conceived the presented idea. E.G. and M.S. performed the computations.  M.S. and A.K.C. wrote the final manuscript.

%\end{acknowledgements}

%
%\bibliographystyle{spphys}
%\bibliography{vaccination}

%\begin{thebibliography}{99}
%\end{thebibliography}

% Figures

%fig 1

%fig 2

\clearpage

%fig 3

\clearpage

%fig 4

\clearpage

%fig 5
\begin{figure}[!h]
\begin{center}
\includegraphics[width=0.98\textwidth]{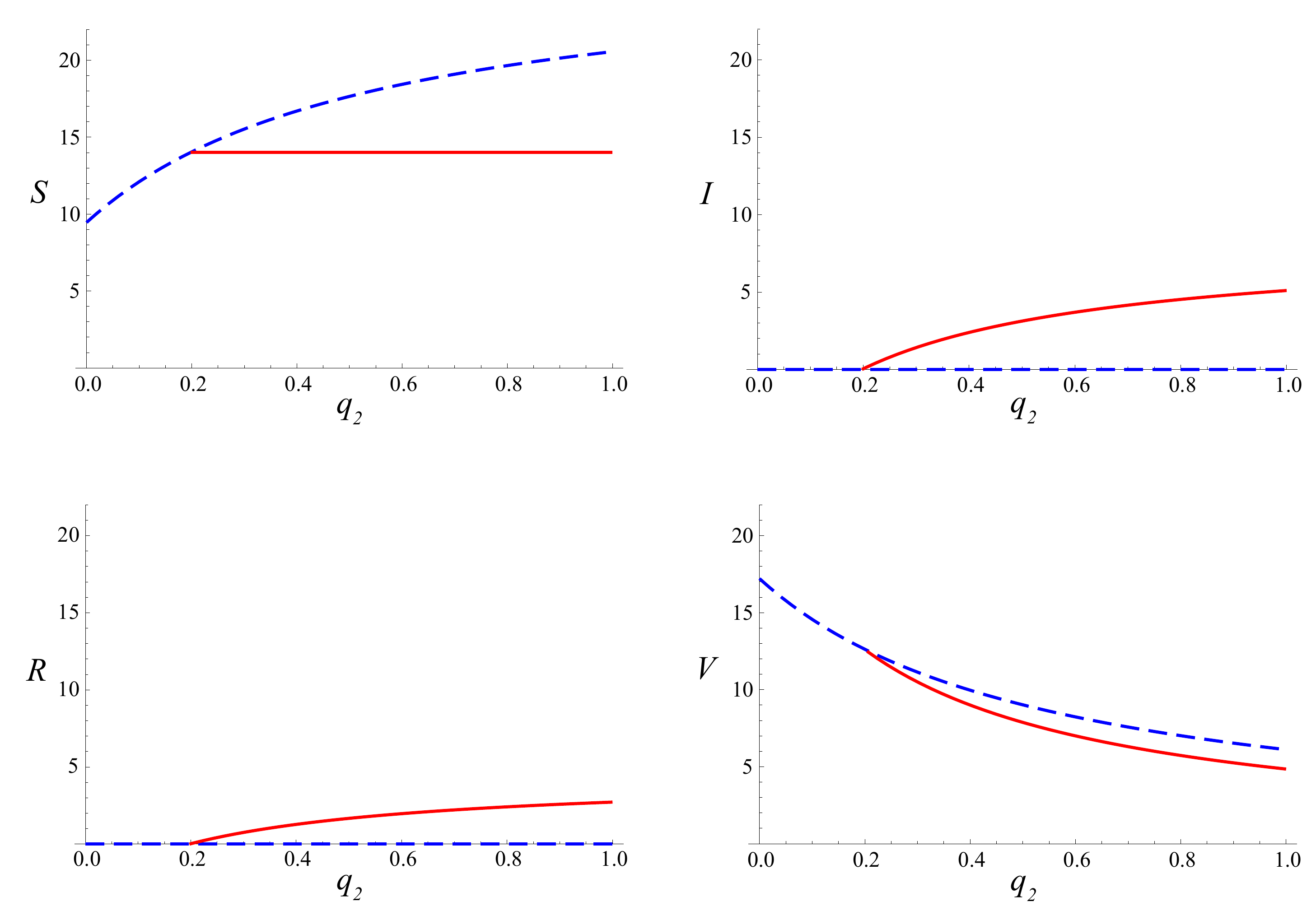}
\end{center}
\caption{The stationary states as function of $q_2$: the red (solid) curve indicates the endemic state, the blue (dashed) curve the disease-free state. Parameter values chosen are as follows: $\beta = 0.05$, $a = 0.4$, $p = 0.3$, $B = 8.0$, $v_1 = 0.35$, $v_2 = 0.25$, $q_1 = 0.45$.}
\label{fig:bifq2}
\end{figure}

\clearpage

%fig 6
\begin{figure}[!h]
\begin{center}
\includegraphics[width=0.98\textwidth]{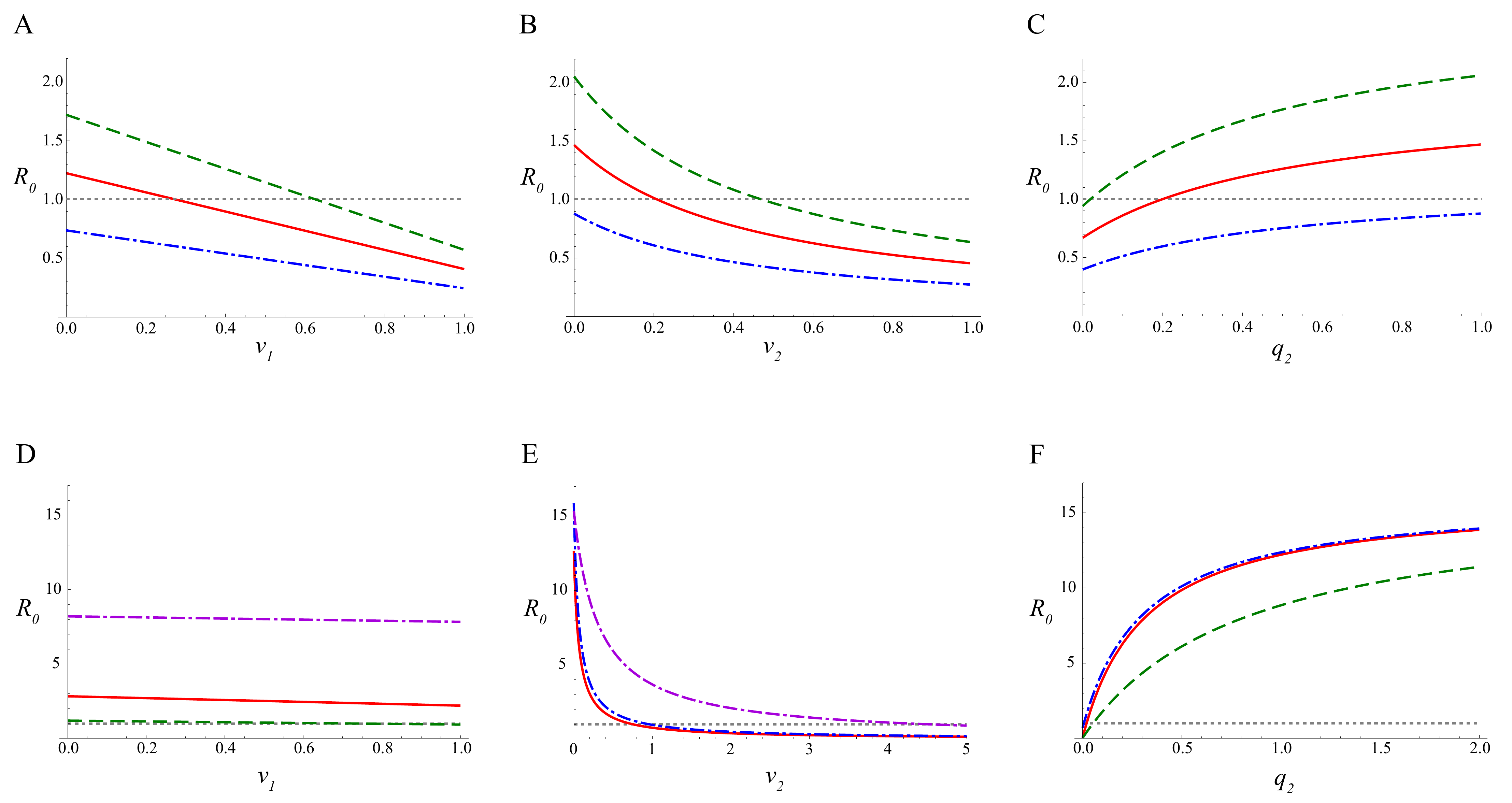}
\end{center}
\caption{(A-C) The basic reproduction number $R_0$ (which is not the same as the initial condition of the number of recovered agents in a simulation) as function of the parameters $v_1$ (A), $v_2$ (B) and $q_2$ (C), and each for three values of the infection rate, $\beta=0.03$ (blue), $\beta=0.05$ (red) and $\beta=0.07$ (green). The curve $R_0=1$ is shown as a thin grey dotted curve. Compare also with Figs.~\ref{fig:bifv1}-\ref{fig:bifq2}.  Other parameter values are as follows: $a = 0.4$, $p = 0.3$, $B = 8.0$, $v_1 = 0.35$, $v_2 = 0.25$, $q_1 = 0.45$, $q_2= 0.15$. (D-F) The basic reproduction number $R_0$ as function of the parameters $v_1$ (D), $v_2$ (E) and $q_2$ (F) for a parameter set describing pertussis infection (parameters taken partially from~\cite{RozhnovaJRSI12} and chosen such that an uncontrolled $R_0$ (no vaccination) is around 16). 
In (D), following curves are shown: $v_2=0.3$, $q_2=0.05$ (red solid curve), $v_2=0.8$, $q_2=0.05$ (green dashed curve) and $v_2=0.3$, $q_2=0.3$ (purple dot-dashed curve). In (E), following curves are shown: $v_1=0.95$, $q_2=0.05$ (red solid curve), $v_1=0.05$, $q_2=0.05$ (blue dot-dashed curve) and $v_1=0.95$, $q_2=0.3$ (purple dot-dashed curve). In (F), following curves are shown: $v_1=0.95$, $v_2=0.3$ (red solid curve), $v_1=0.95$, $v_2=0.8$ (green dashed curve) and $v_1=0.05$, $v_2=0.3$ (blue dot-dashed curve). The other parameters are $\beta = 140$, $p = 1/70$, $B = 0.026$, $q_1 = 0.1$ (assuming that time is measured in years) and the curve $R_0=1$ is shown as a thin grey dotted curve.}
\label{fig:R0}
\end{figure}

\clearpage

%fig 7

\clearpage

%fig 8

\clearpage

%fig 9

\clearpage

%fig 10

\end{document}